\documentclass[%
 reprint,
 amsmath,amssymb,
 aps,prb
]{revtex4-1}
\usepackage{graphicx}% Include figure files
\usepackage{dcolumn}% Align table columns on decimal point
\usepackage{bm}% bold math
\usepackage{xcolor}
\begin{document}
\title{Giant nonlinear response due to unconventional oscillation in Nodal-line semimetals}
\author{Debabrata Sinha$^{1,2}$ and A. Taraphder$^1$}
\affiliation{$^1$Department of Physics, Indian Institute of Technology, Kharagpur-721302, India\\$^2$Institute of Mathematical Sciences, Taramani, Chennai 600113, India}

\date{\today}% It is always \today, today,

\begin{abstract}

Quantum oscillations in magnetoconductance of a material at low temperatures and in presence of an intense magnetic field are described by the Shubnikov de Haas (SdH) effect. It is widely assumed to be the hallmark of the Fermi surface of a given metal. In contrast to the canonical situation, we identify an exotic oscillation in nonlinear responses of three-dimensional nodal line semimetals (NLSMs) which persist even at temperatures where the typical SdH-like oscillations vanish. This oscillation occurs due to the periodic gap-closing of a pair of Landau levels at zero Fermi energy with the variation of the magnetic field. The emergence of the oscillation is a remarkable fingerprint of ring dispersion and the corresponding frequency can be used to determine the radius of the ring. Using the Boltzmann equation, we calculate the second harmonic generation of nodal line semimetals under parallel DC electric and strong magnetic fields. The second harmonic conductivity diverges at the gap closing condition leading to the giant nonlinear response in NLSMs.

\end{abstract}
\maketitle

\textbf{Introduction:-} The interaction between solids and laser field induces innumerable nonlinear effects in a material\cite{Bolem-Non,Bai-Nat21,Wu-Nat17,Juan-NatCom17}. Among these, the second harmonic nonlinear effect occurs in a crystal without inversion symmetry. The nonlinear response became a basic tool to examine the various electronics properties such as dynamical Bloch oscillation\cite{Schubert-Nat}, quantum interference\cite{Hohenleutner-Nat} and band geometry\cite{Fu-PRL15, Moore-PRL10} of a given crystal. In recent years, the nonlinear effects of three-dimensional topological materials like Dirac and Weyl semimetals have drawn much attention\cite{Takasan-20}. The effects in those materials are found to be several times larger than the semiconductors or other conventional metals and yield potential applications for optical devices. Experimentally, a giant second-order nonlinear effect has been reported in inversion broken Weyl semimetals (WSMs) such as TaAs, TaP, and NbAs\cite{Wu-Nat17}. A large value of second-order nonlinear susceptibility $\chi^{(2)}$ predicted in Dirac semimetals and magnetic WSMs\cite{Takasan-20,Gao-PRB21}. Most of the intriguing nonlinear phenomena including quantized circular photogalvanic effect\cite{Juan-NatCom17, Flicker-PRB18}, shift current\cite{Gavin-Nat19}, photocurrent\cite{Chan-PRB17} in three dimensional (3D) topological materials are stemmed from their low energy band structure. 

Nodal line semimetals (NLSMs) are another class of 3D topological materials having linear dispersion around a one dimensional (1D) loop in $k$-space\cite{Fang-PRB15, Burkov-PRB11}. These materials are distinct from other 3D topological materials like Weyl or Dirac in which the linear dispersion occurs around some discrete points in the Brillouin zone. Many nodal line materials have been found experimentally, like HgCr$_2$Se$_4$\cite{Xu-PRL11}, Cu$_3$(Pd/Zn)N\cite{Yu-PRL15,Kim-PRL15}, SrIrO$_3$\cite{Chen-Nat15}, Ca$_3$P$_2$\cite{Chan-PRB16} and ZrSiS\cite{Schoop-Nat16}. The experimental characterization of their nodal loop structure can be probed by quantum oscillations (QOs) in presence of an external magnetic field\cite{Oro-PRB18,Yang-PRB18,Li-PRL18,Kar-JPCM21}. The oscillation in resistivity, known as SdH oscillation, originated from the periodic crossing of quantized Landau levels(LLs) with chemical potential. The oscillation period is related to the shape of the Fermi surface. In addition to this, in NLSMs, a pair of LLs periodically intersects each other at zero Fermi energy with a variation of the magnetic which violates the conventional QOs theory. This novel oscillation is a remarkable fingerprint of nodal line dispersion and it is absent in other topological materials like Dirac and Weyl semimetals. We ask the question: whether the exotic oscillation leads to any novel and technologically useful nonlinear properties in NLSM. Moreover, can there be similar SdH oscillation in nonlinear response which can probe the Fermi surface of NLSMs. This is because the hallmark of NLSM in nonlinear response is still missing.

In this Letter, we study the nonlinear responses in NLSM in presence of a strong magnetic field, using the semiclassical Boltzmann transport equation. A small DC electric field is introduced to break the inversion symmetry. We calculate the current induced second harmonic generation (SHG). The second harmonic conductivity (SHC) is found to exhibit oscillations with inverse magnetic field $1/B$ at temperature $T\rightarrow 0$. The oscillation is akin to the SdH oscillation in which extremal orbits are identified through the magnetic field dependence. At finite temperature, SHC diverges periodically with a variation of the inverse magnetic field $1/B$. The divergence occurs at the specific values of magnetic fields where a pair of LLs closes their gap at zero Fermi energy. The second harmonic nonlinear susceptibility $\chi(2\omega)$ has the largest value around those specific values of magnetic fields. We also predict this novel oscillation exists in higher harmonic generations of NLSMs.

\textbf{Model Hamiltonian and LL Spectrum:-} Let us consider a 3D NLSM Hamiltonian in low energy, is given by,\cite{Yang-PRB18,Cortijo-PRB18,Rui-PRB18},
\begin{eqnarray}
\mathcal{H}=((p^2_x+p^2_y)/2m^*-\Delta)\sigma_x+vp_z\sigma_z
\label{mod-hamil}
\end{eqnarray}
where $\sigma$ is the Pauli matrices acting on the orbital space, $m^*>0$ is the effective mass, $v$ is the Fermi velocity in the $z$-direction and $\Delta>0$ is a constant in energy scale. The model Hamiltonian in Eq.(\ref{mod-hamil}) preserves both parity $\mathcal{P}=\sigma_x \otimes (\mathbf{k} \rightarrow -\mathbf{k})$ and time-eversal symmetry $\mathcal{T}=K\otimes (\mathbf{k} \rightarrow -\mathbf{k})$, where $K$ is the complex conjugation. The energy spectrum of the Hamiltonian is given by,
\begin{eqnarray}
E_{\pm}=\pm\sqrt{(\Delta-(p^2_x+p^2_y)/2m^*]^2+v^2p^2_z}
\label{nodal-spec}
\end{eqnarray}
The upper and lower energy bands touches along a circle defined by $p^2_0=p^2_x+p^2_y=2m^*\Delta$, at $p_z=0$ plane in BZ with a radius $\sqrt{2m^*\Delta}$. At low energy, the system Hamiltonian yields the Dirac cone at $\pm p_0$. The Fermi surface takes the shape of torus of genus one, for $\mu< \Delta$, where $\mu$ is the chemical potential. The Fermi surface has a drumlike structure for $\mu>\Delta$. Further increase of $\mu$, the surface geometry changes to sphere of genus zero.
\begin{figure}
\includegraphics[scale=.36]{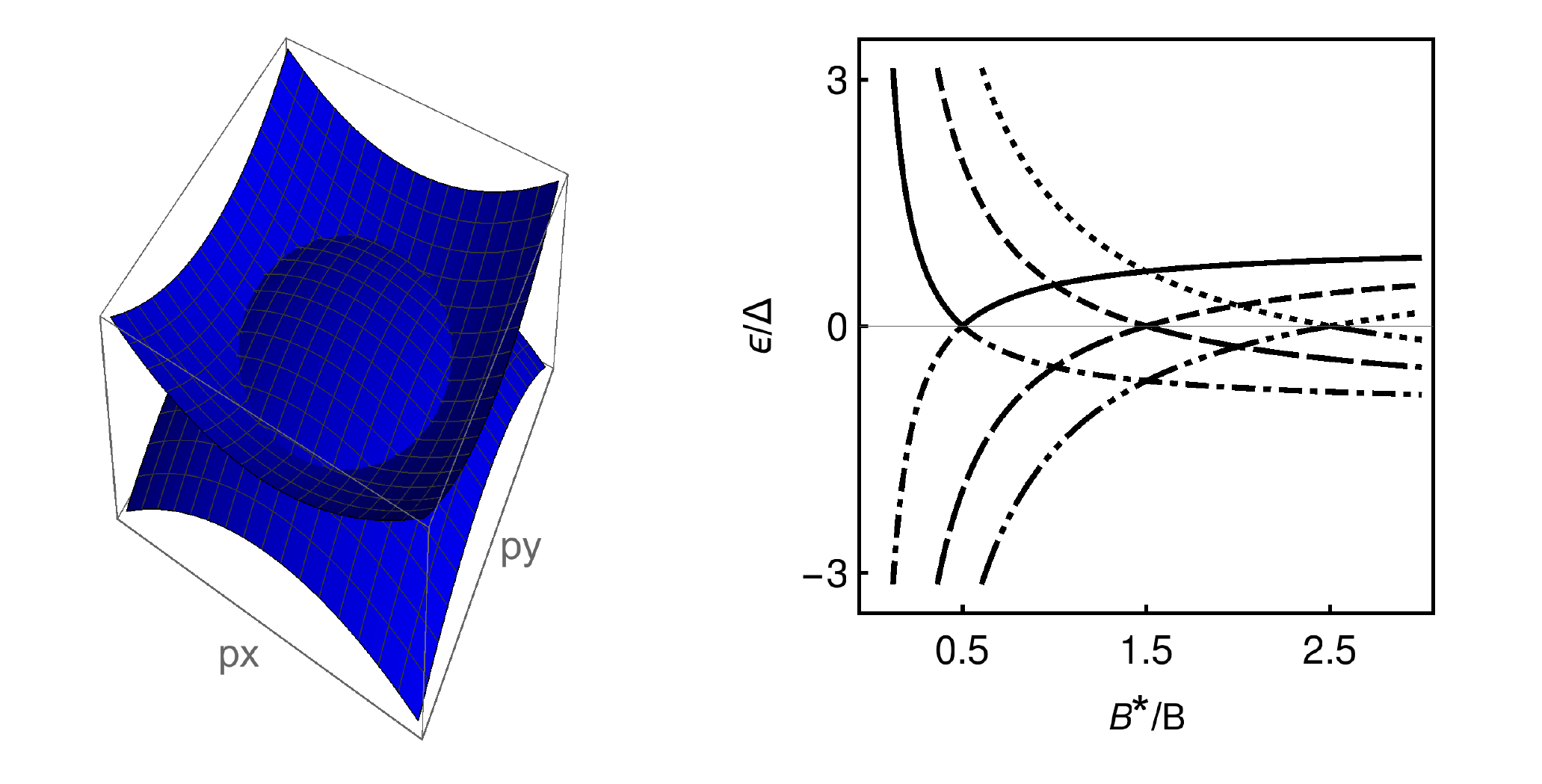}
\includegraphics[scale=.385]{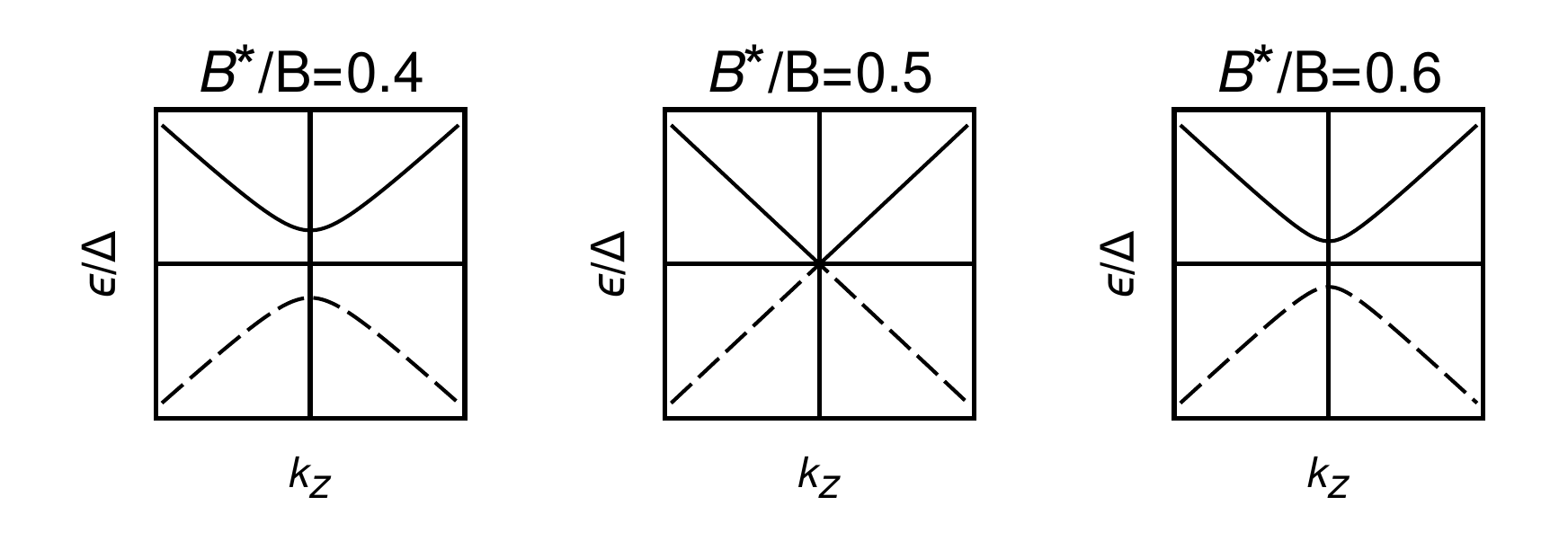}
\caption{Upper panel displays the zero magnetic field spectrum (in the left). Right hand side, we show the Landau level spectrum with inverse magnetic field. The crossing point of a pair of Landau Levels appear periodically with $1/B$. Lower panel displays $n=0$ Landau level spectrum with increasing $1/B$ field. The Landau levels simplies to a linear spectrum for $B^*/B=0.5$.}
\end{figure}
We apply the magnetic field along $z$-direction i.e., $\mathbf{B}=(0,0,B)$ and choose the vector potential is $\mathbf{A}=(-yB,0,0)$. Define ladder operators, $a\equiv -[(y-l^2_Bk_x)/l_B+l_B \partial_y]/\sqrt{2}$ and $a^\dagger\equiv -[(y-l^2_Bk_x)/l_B-l_B \partial_y]/\sqrt{2}$ with magnetic length $l_B=\sqrt{\hbar/eB}$. We can rewrite the Hamiltonian in Eq.(\ref{mod-hamil}) in terms of Ladder operators as follows,
\begin{eqnarray}
\mathcal{H}=[\frac{\hbar e B}{m_*}(a^\dagger a+1/2)-\Delta]\sigma_x+v p_z \sigma_z
\label{Landau-hamil}
\end{eqnarray}
The Landau eigenenergies of Eq.(\ref{Landau-hamil}) are given by\cite{Yang-PRB18,Li-PRL18},
\begin{eqnarray}
\epsilon^\pm_n=\pm\sqrt{[\Delta-\frac{\hbar e B}{m_*}(n+1/2)]^2+v^2p^2_z}
\label{landau-spec}
\end{eqnarray}
where $n=0,1,2....$ is the Landau index. The pairs of Landau Levels $\epsilon^+_n$ and $\epsilon^-_n$ for a given $n$, crosses each others with the variation of magnetic field. The crossing occurs at a set of magnetic fields,
\begin{eqnarray}
B_n=\frac{m^*\Delta}{e\hbar (n+1/2)}
\label{mag_cross}
\end{eqnarray}
In the rest of the paper we scale energy $\epsilon_n$ by $\Delta$ and magnetic field $B$ by $B^*=m^*\Delta/e\hbar$.
\begin{figure}
\includegraphics[scale=.45]{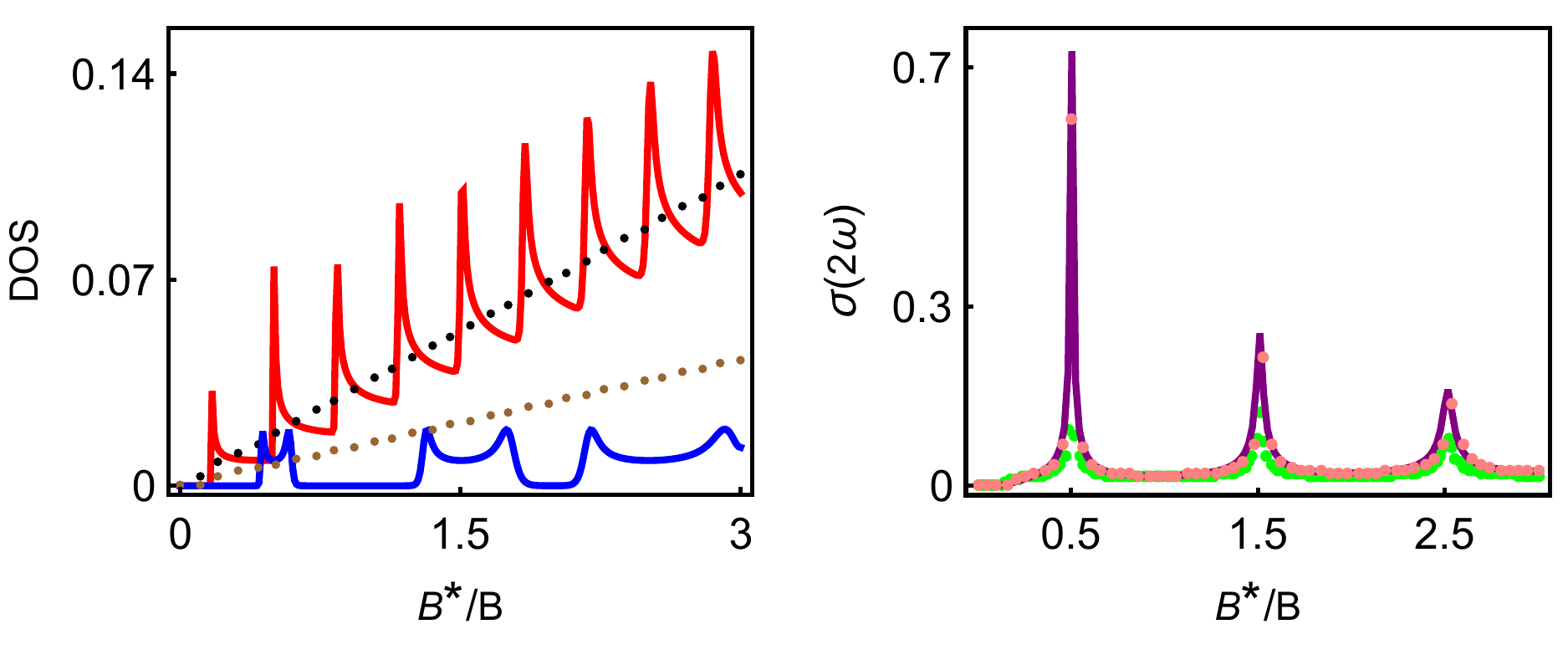}
\caption{Left Panel, displays the density of states (DOSs), in Eq.(\ref{dos}) of main text, with inverse magnetic field. The red and blue solid lines correspond to $\mu/\Delta=2$ and $0.1$, respectively. The temperature is fixed at $k_BT/\Delta=.01$. The dotted lines are the corresponding classical natures at finite temperature, $k_BT/\Delta=1$. Right panel, shows the second order conductivity $\sigma(2\omega)$ (in unit of $e^3\tau/\hbar^2$) with inverse magnetic field for different $\mu$. The temperature is fixed at $k_BT/\Delta=1$, where the oscillations in DOS vanishes. The conductivity $\sigma(2\omega)$ diverges at the quantized value of $B^*/B=(n+1/2)$, with $B^*=m^*\Delta/e\hbar$. Here, we take a finite value of mass $m/\Delta=0.01$, to regularize the divergence in the peack. The purple solid line, green dotted line and black dashed line corresponds to $\mu/\Delta=.5$, $1$ and $2$, respectively. See text for details.}
\label{main_fig}
\end{figure}

\textbf{Current induced SHG:-} We calculate the response tensor of current induced second harmonic generation by using Boltzman transport equation. A weak DC electric field $\mathbf{E}_{dc}$ is introduced which lead the system into a nonequllibrium steady state. In the relaxation time approximation, the steady state Boltzmann equation for the electron distribution $f^n$ in the $n$-th Landau level is given by,
\begin{eqnarray}
-\frac{e\mathbf{E}_{dc}}{\hbar}\cdot \nabla_\mathbf{k}f^n_{dc}=-\frac{f^n_{dc}-f^n_0}{\tau}
\label{Boltz-dc}
\end{eqnarray}
where $f^n_0=(e^{(\epsilon_n-\mu)/k_BT} +1)^{-1}$, is the equilibrium distribution function. Here, $k_B$ is the Boltzman constant, $T$ is temperature and $\tau$ is the relaxation time. The solutuion with linear $\mathbf{E}_{dc}$ of the above equation is given: $f^n_{dc}=f^n_0+\frac{e\tau \mathbf{E}_{dc}}{\hbar}\cdot\nabla_\mathbf{k}f^n_0$. Thus the DC electric field globally shift the Fermi surface by $\frac{e\tau \mathbf{E}_{dc}}{\hbar}$ in the $\mathbf{k}$-th direction. We now apply a strong laser beam of frequency $\omega$ and electric field amplitude $\mathbf{E}_{o}$. The total electric field is in the form $\mathbf{E}_{tot}=\mathbf{E}_{dc}+E_{op}(t)=\mathbf{E}_{dc}+\mathbf{E}_{0}e^{-i\omega t}+c.c$.
The Boltzman equation in presence of both DC and optical field $E_{op}(t)$ is given by,
\begin{eqnarray}
{\partial f^n_{op}\over \partial t}-\frac{e(\mathbf{E}_{dc}+\mathbf{E}_{op}(t))}{\hbar}\cdot\nabla_\mathbf{k} f^n_{op}-\frac{e(\mathbf{E}_{dc}+\mathbf{E}_{op}(t))}{\hbar}\cdot\nabla_{\mathbf{k}} f^n_{dc}\nonumber\\=-\frac{f^n_{dc}+f^n_{op}-f^n_0}{\tau}\nonumber\\
\end{eqnarray}
Using Eq.(\ref{Boltz-dc}) and considering weak DC field, the above equation becomes\cite{Gao-PRB21,Cheng-Opt14},
\begin{eqnarray}
{\partial f^n_{op}\over \partial t}-\frac{e\mathbf{E}_{op}(t)}{\hbar}\cdot\nabla_{\mathbf{k}} f^n_{op}-\frac{e\mathbf{E}_{op}(t)}{\hbar}\cdot\nabla_{\mathbf{k}} f^n_{dc}=-\frac{f^n_{op}}{\tau}\nonumber\\
\end{eqnarray}
The distribution function $f^n_{op}$ can be expanded in a power series in the electric field as follows,
\begin{eqnarray}
f^n_{op}&=&f^{(n,\omega)}_{op}e^{-i\omega t}+f^{(n,2\omega)}_{op}e^{-2i\omega t}+...\nonumber\\&&+f^{(n,m\omega)}_{op}e^{-mi\omega t}\nonumber+cc+..
\end{eqnarray}
where $f^{(n,m\omega)}_{op}\propto \mathbf{E}^m_{op}$ is the $m$-th order corrections of the distribution function. It the following form,
\begin{eqnarray}
\label{first-order-dist}
f^{(n,\omega)}_{op}&=&\frac{e\tau \mathbf{E}_{op}}{\hbar(1-i\omega \tau)}\cdot\nabla_{\mathbf{k}} f^n_{dc}\nonumber\\
f^{(n,2\omega)}_{op}&=&\frac{(e\tau)^2 \mathbf{E}^2_{op}}{\hbar^2(1-i\omega \tau)(1-2i\omega \tau)}\nabla^2_{\mathbf{k}} f^n_{dc}\nonumber\\
.\nonumber\\
.\nonumber\\
f^{(n,m\omega)}_{op}&=&\frac{(e\tau)^m \mathbf{E}^m_{op}}{\hbar^m(1-i\omega \tau)(1-2i\omega \tau)...(1-mi\omega \tau)}\nabla^m_{\mathbf{k}} f^n_{dc}\nonumber\\
\label{second-order-distr}
\end{eqnarray}
We consider the electric fields $\mathbf{E}_{dc}$ and $\mathbf{E}_{op}$ are along the $z$-direction i.e., $\mathbf{E}_{dc}\parallel \mathbf{B}$ and $\mathbf{E}_{op}\parallel \mathbf{B}$. So, we find only the $z$-cmponent of charge density and $zzz$ component of nonlinear conductivity tensor. The charge current density of the $m$-th order is given by,
\begin{eqnarray}
j^{(n,m\omega)}_z=\frac{-e}{2\pi l^2_B}\sum_n\int v_{z,n}f^{(n,m\omega)}\frac{dk_z}{2\pi}
\label{current-dens}
\end{eqnarray}
We obtain the linear response by using the distribution function in Eq.(\ref{first-order-dist}). Similarly, the second order response is derived by using Eq.(\ref{second-order-distr}) and the SHG current density is explicitly written as\cite{Gao-PRB21,Cheng-Opt14},
\begin{eqnarray}
j^{(n,2\omega)}_z=\frac{-e}{2\pi l^2_B}\frac{(e\tau \mathbf{E}_{op})^2}{\hbar^2(1-2i\omega \tau)(1-i\omega \tau)}\nonumber\\
\sum_n\int v_{z,n}\nabla^2_{\mathbf{k}}(f^n_0+\frac{e\tau E_{dc}}{\hbar}{\partial f^n_0 \over \partial k_z})\frac{dk_z}{2\pi}
\end{eqnarray}
The first term in the integration $v_{l,n}\nabla^2_{\mathbf{k}}f^n_0$ is zero since it is odd under $\mathbf{k}\rightarrow -\mathbf{k}$. Thus, the final form of SHG current density is given by,
\begin{eqnarray}
j^{(n,2\omega)}_z=\frac{-e^4\tau^3 \mathbf{E}^2_{op} E_{dc}}{(2\pi l^2_B)\hbar^3(1-2i\omega \tau)(1-i\omega \tau)}\nonumber\\
\sum_n\int v_{z,n}{\partial^3 \epsilon_n \over \partial k^3_z}{\partial f^n_0 \over \partial \epsilon_n}\frac{dk_z}{2\pi}
\label{SHC}
\end{eqnarray}
Using the relation $j^{(n,2\omega)}_l=\sigma_{n,l}(2\omega)\mathbf{E}^2_{op}$, we obtained the second harmonic conductivity tensor $\sigma_{n,l}(2\omega)$:
\begin{eqnarray}
\sigma(2\omega)=\frac{\sigma_{0z}}{(1-2i\omega \tau)(1-i\omega \tau)}
\end{eqnarray}
where,
\begin{eqnarray}
\sigma_{0z}=\frac{1}{4\pi^2}\frac{B}{B^*}\frac{E_{dc}}{E^*_{dc}}\sum_n\int v_{z,n}{\partial^3 \epsilon_n \over \partial k^3_z}{\partial f^n_0 \over \partial \epsilon_n}dk_z
\label{cond_landsub}
\end{eqnarray}
Here, the conductivity $\sigma_{0z}$ is scled by $e^3\tau/\hbar^2$ and $E^*_{dc}=\hbar \Delta/e \tau^2 v^3 m^*$. The behaviour of quantum transport is controlled by the derivative of the Fermi function, which is given by,
\begin{eqnarray}
{\partial f^n_0 \over \partial \epsilon}=-\frac{1}{2k_BT}\frac{1}{1+\cosh[\frac{(\epsilon_n-\mu)}{k_BT}]}
\label{fermi_temp}
\end{eqnarray}
which has a peak at $\epsilon_n=\mu$ with a width depends on temperature. We define the the density of states near the Fermi level\cite{Knolle-PRL17,Zhang-PRL16},
\begin{eqnarray}
DOS=\frac{1}{2k_BT}\sum_n\frac{1}{1+\cosh[\frac{(\epsilon_n-\mu)}{k_BT}]}
\label{dos}
\end{eqnarray}

\begin{figure}
\includegraphics[scale=.45]{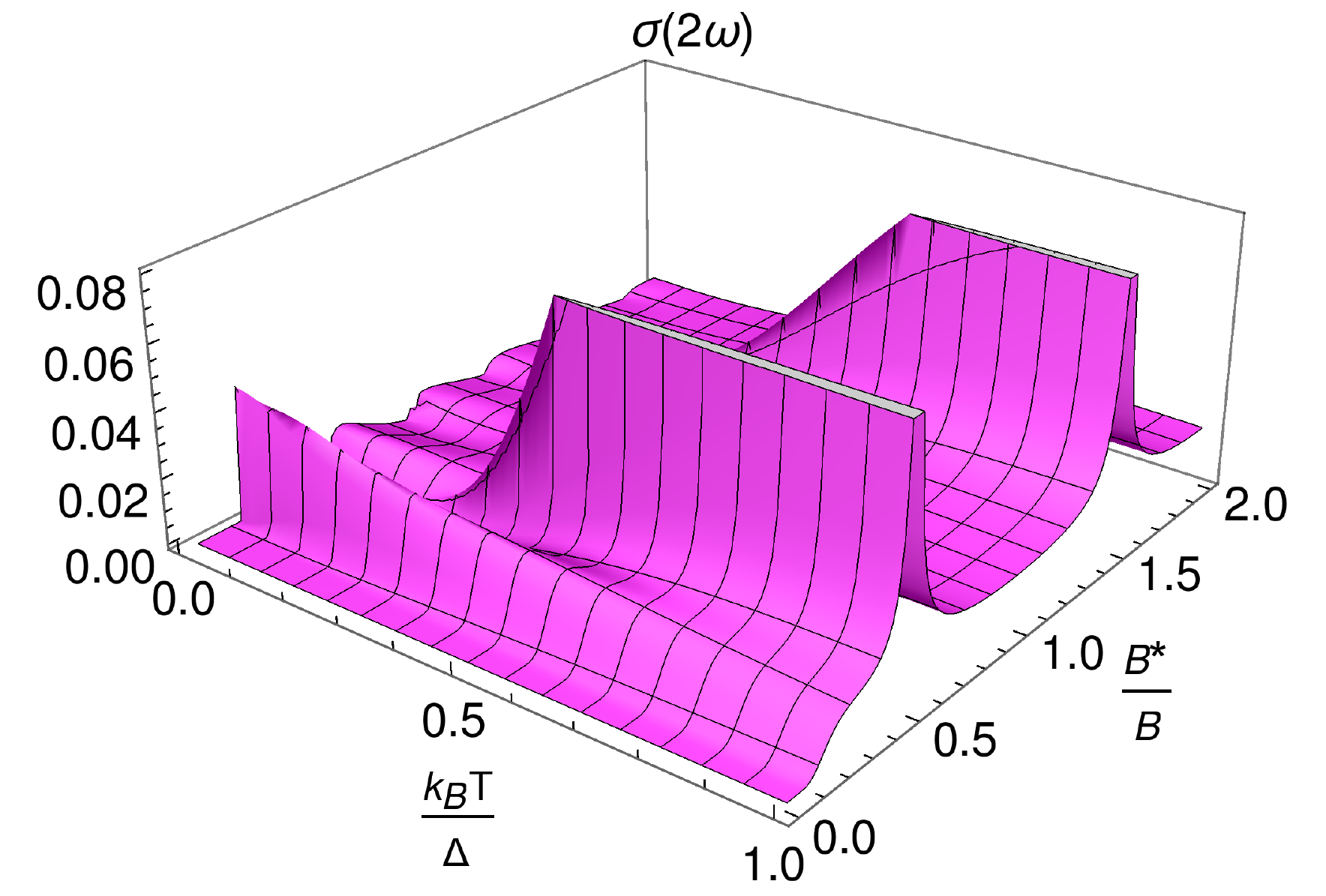}
\caption{The evolution of current induced SHC in NLSMs as a function of temperature and inverse magnetic field.}
\label{3d_plot}
\end{figure}

\textbf{Results and Discussions:-} The left panel of Fig.(\ref{main_fig}) displays the variation of DOS as a function of inverse magnetic field for two different chemical potentials. At sufficiently low tmperature $k_BT<<\hbar \omega_B=\hbar e B/m^*$, there are $1/B$-periodic quantum oscillation in DOS. We find the periodicity in $1/B$ is $(1+\mu/\Delta)^{-1}/B^*$ (shown by red solid line) for $\mu>\Delta$. On the other hand,  for $\mu<\Delta$, the oscillation appears with two peaks. The peacks have different periodicities in $1/B$, which are $(1+\mu/\Delta)^{-1}/B^*$ and $(1-\mu/\Delta)^{-1}/B^*$, respectively (shown by blue solid line). The peak of the oscillation occurs in each time when a Landau level crosses the chemical potential. The double peaks are associated with two extremal Fermi surfaces of NLSMs for $\mu<\Delta$. As the temeprature increased, the oscillation corresponding to Landau level gradually smoothed out as shown by black and brown dotted lines. When $k_BT>>\hbar\omega_B$, the integration factor in Eq.(\ref{cond_landsub}) simplifies,
\begin{eqnarray}
\sum_n\int v_{z,n}{\partial^3 \epsilon_n \over \partial k^3_z}{\partial f^n_0 \over \partial \epsilon_n}dk_z \simeq \frac{\pi}{64k_BT}\sum_n\frac{Sech^2(\frac{|\mathcal{B}_n|-\mu}{2k_B T})}{|\mathcal{B}_n|}\nonumber\\
\label{sing_cond}
\end{eqnarray}
where $\mathcal{B}_n=(1-B(n+1/2)/B^*)$. The right hand side function in Eq.(\ref{sing_cond}) has singularities at $\mathcal{B}_n\rightarrow 0$ i.e., $B^*/B=(n+1/2)$. As a result, the conductivity tensor $\sigma(2\omega)$ diverges periodically with an inverse magnetic field with periodicity $1/B=e\hbar/m^*\Delta$. This is one of the main results of our work. We have shown this in the right panel of Fig.(\ref{main_fig}). We consider a $\mathcal{P}\mathcal{T}$ breaking mass term ($m\sigma_y$) in the Hamiltonian in Eq.(\ref{mod-hamil}) to regularize the divergence\cite{Cortijo-PRB18,Rui-PRB18}. The right hand side of Eq.(\ref{sing_cond}) now modifies to:
\begin{eqnarray}\frac{\pi}{64k_BT}\sum_n\frac{Sech^2(\frac{\sqrt{m^2+\mathcal{B}^2_n}-\mu}{2k_BT})}{\sqrt{m^2+\mathcal{B}^2_n}}\end{eqnarray}. This quantity diverges at $m\rightarrow 0$ and $\mathcal{B}_n\rightarrow 0$, which appears periodically with $1/B$. The frequency of this oscillation is $\mu$ independent.

From Eq.(\ref{mag_cross}), the LLs of given index $n$ simplifies to a linear spectrum at the specific values of magnetic field: $B^*/B_n=(n+1/2)$. The pair of $n$-th LLs intersect each other at zero Fermi energy and the gap vanishes. Consequently, the conductivity $\sigma_n(2\omega)$ and higher-order conductivities will vanish since they involve higher-order derivatives (see Eq.(\ref{SHC})). The linear spectrum occurs for each Landau index $n$ which gives rise to a $\mu$-independent oscillation. This is in contrast to Weyl or Dirac semimetals where the linear dispersion only appears for $n=0$\cite{Deng-PRL19}. However, this oscillation gets suppressed by the SdH oscillation, shown in Fig.(\ref{3d_plot}). This is because at Low temperature, Eq.(\ref{fermi_temp}) becoming a delta function, signifies that only near the Fermi surface electron's contributes to the SHG. As a result, the function in Eq.(\ref{sing_cond}) does not diverge and only SdH oscillation persists. With increasing the temperature, electrons deep in the Fermi sphere are also taken into the process of SHG. That makes the function in Eq.(\ref{sing_cond}) divergent at $\mathcal{B}_n \rightarrow 0$ and at finite temperature. 

\begin{figure}
\includegraphics[scale=.42]{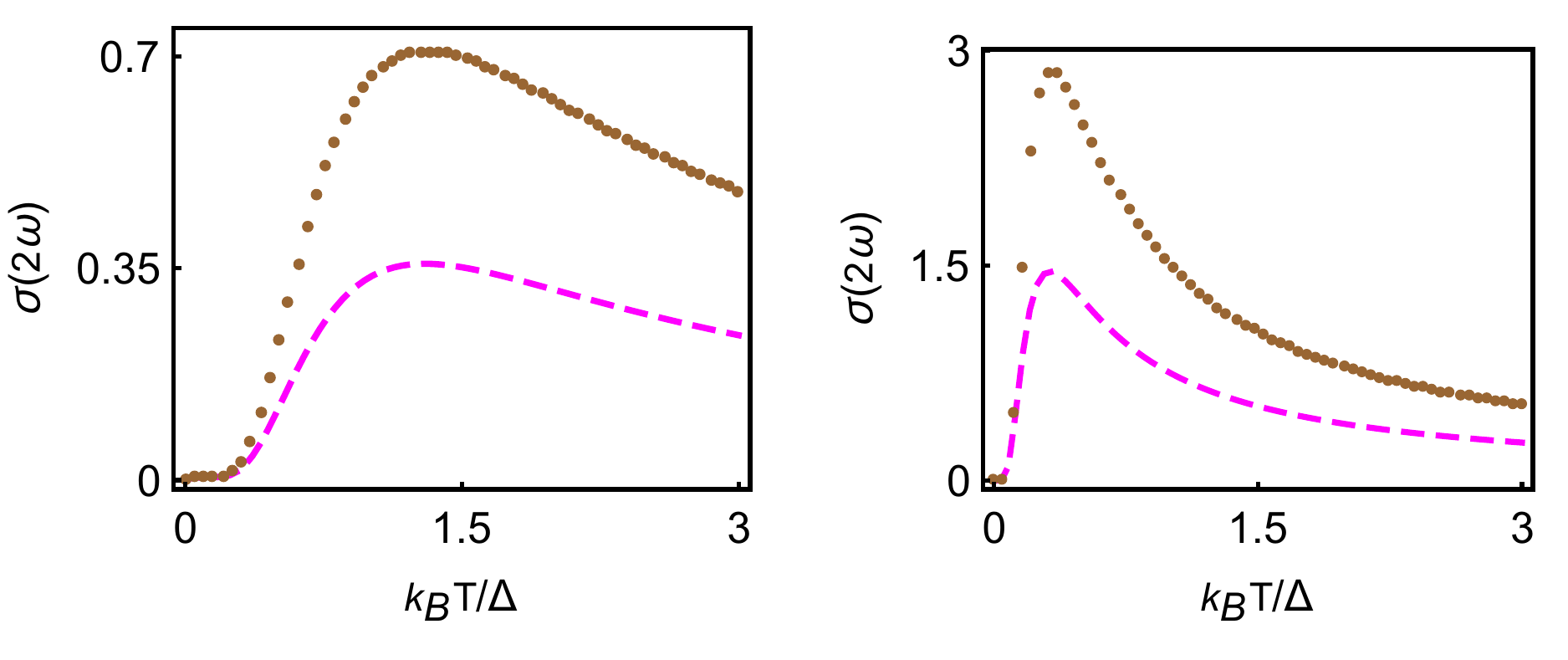}
\caption{\textbf{Temperature Dependence:} $\sigma(2\omega)$ is plotted against temperature $k_BT/\Delta$ for $\mu>\Delta$ (left panel) ($\mu/\Delta=2$) and $\mu<\Delta$ (right panel) ($\mu/\Delta=0.5$). In both panels the brown dotted line and magneta solid line correspond to $m/\Delta=.005$ and $.01$, respectively. We fixed the value $B^*/B=0.5$ in both panels.}
\label{temp_plot}
\end{figure}

We show the temperature dependence of $\sigma(2\omega)$ in Fig.(\ref{temp_plot}), with different mass values $m$ and for a fixed value of $B^*/B=0.5$. The amplitudes of $\sigma(2\omega)$ increases with temperature upto a maximum value and then decreases. The maximum amplitude of $\sigma(2\omega)$ diverges at $m\rightarrow 0$. The temperature $T_d$ at which $\sigma(2\omega)$ has largest value can be determined from the condition: $\frac{d\sigma_{0z}}{dT}|_{T=T_d}=0$ with $m\rightarrow 0$ and $\mathcal{B}_n\rightarrow 0$. We find, $k_BT_d/\Delta \simeq .64\mu/\Delta$.

\begin{figure}
\includegraphics[scale=.18]{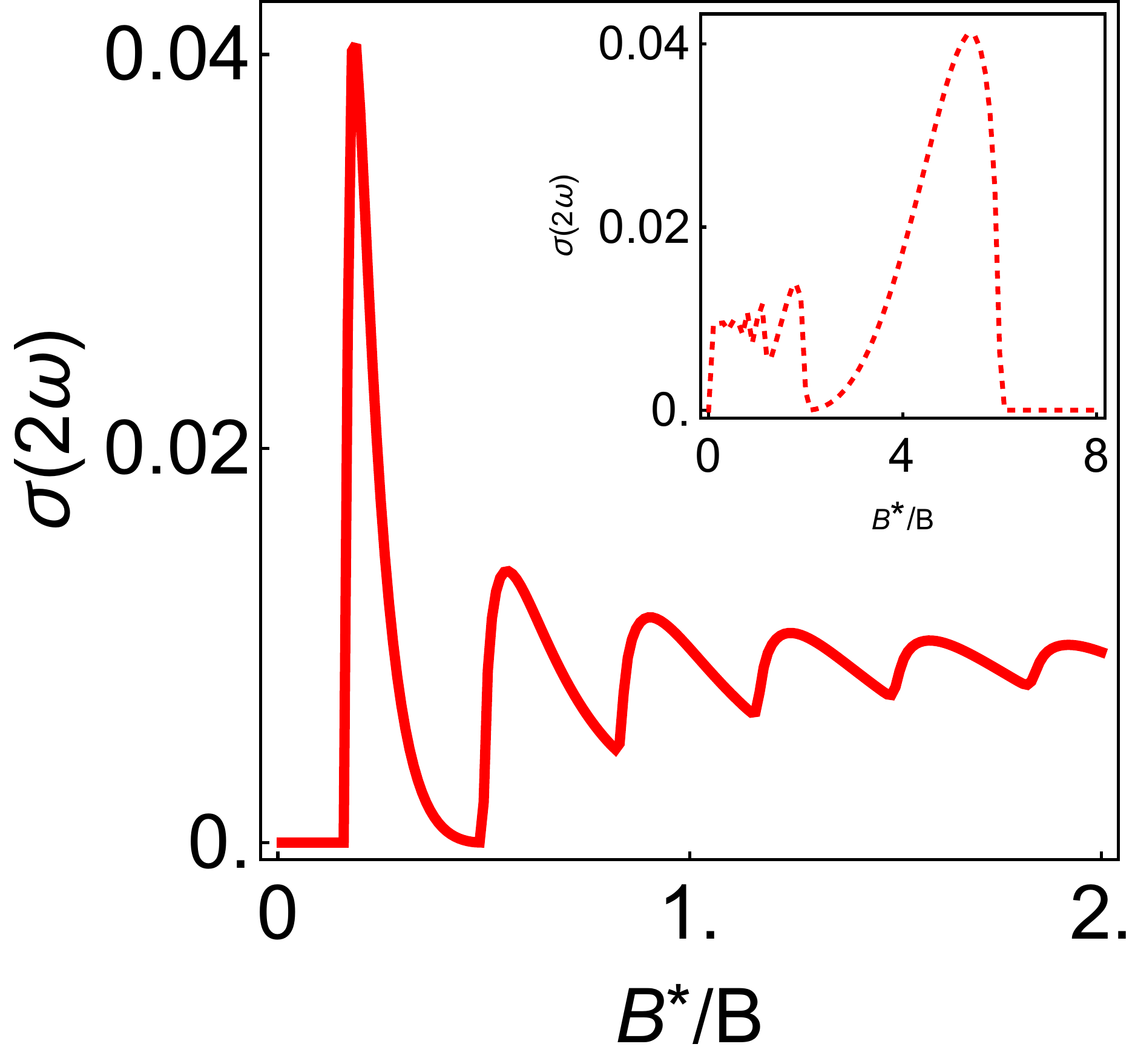}
\includegraphics[scale=.17]{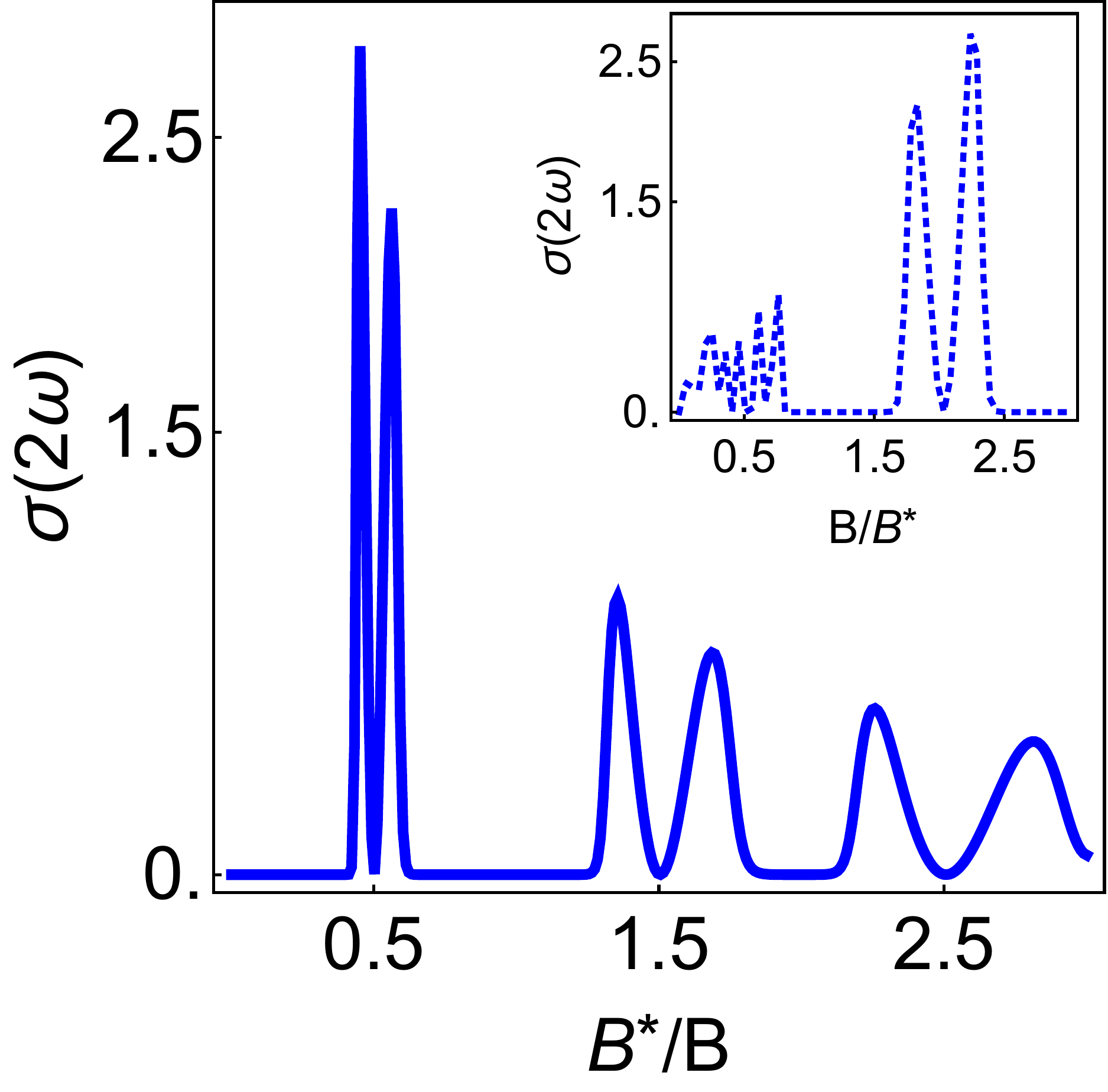}
\caption{The total second harmonic conductivity $\sigma(2\omega)$ as a function of inverse magnetic field at $k_BT/\Delta=0.01$. We fix $\mu/\Delta=2$ and $\mu/\Delta=.15$ in the left and right panel, respectively. Inset shows the variation of $\sigma(2\omega)$ with magnetic field.}
\label{low_temp_sigma}
\end{figure}

\begin{figure}
\includegraphics[scale=.35]{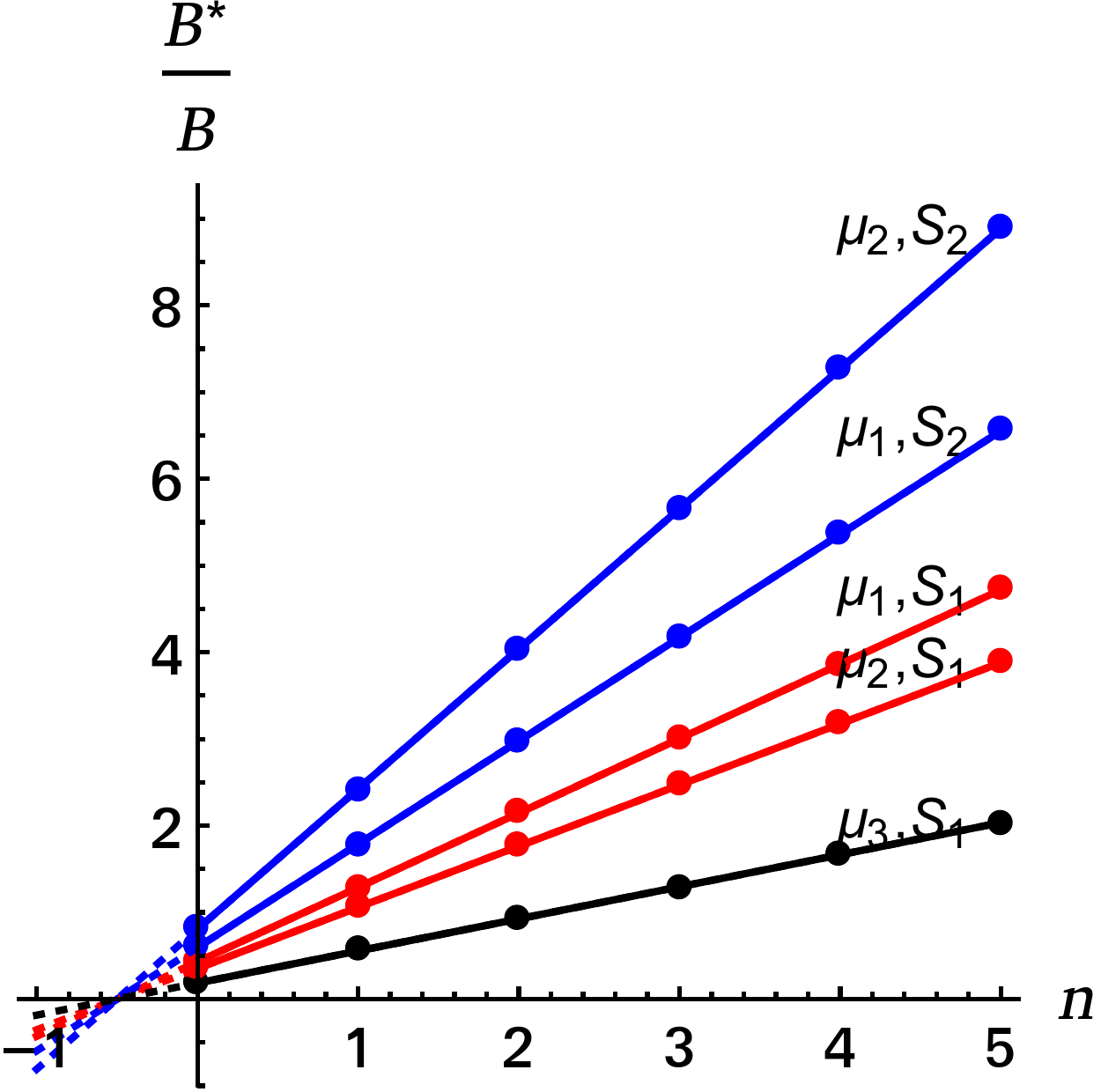}
\includegraphics[scale=0.4]{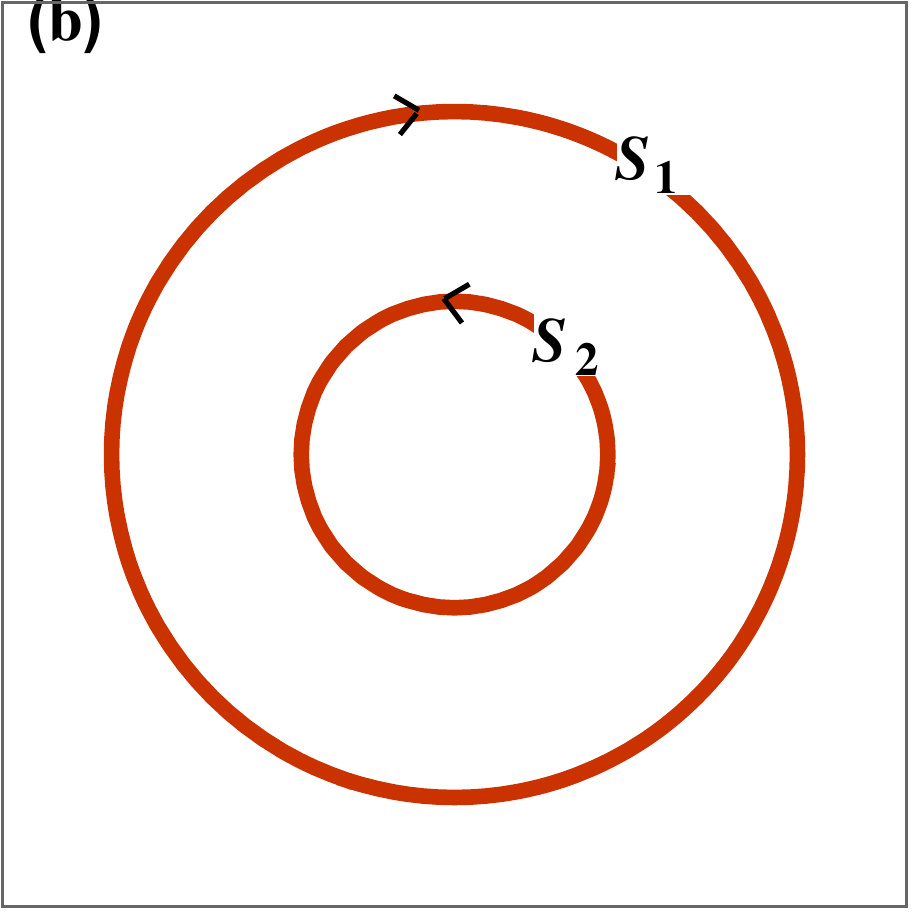}
\caption{Landau index $n$ is plotted against peak position ($B^*/B$) for three different chemical potentials values $\mu_3>\mu_2>\mu_1$ and two corresponding cross sectional area $S_{1,2}$. The value of chemical potentials are $\mu_3/\Delta=2$ (black line), $\mu_2/\Delta=0.5$ (blue and red lines) and $\mu_1/\Delta=0.2$ (blue and red lines). The temperature fixed at $k_BT/\Delta=0.01$. Right panel shows the two surfaces $S_1$ and $S_2$.}
\label{fan_diag}
\end{figure}

We now discuss the origin of quantum oscillations in SHC at Low temperature which are shown in Fig.(\ref{low_temp_sigma}). At low temperature,${\partial f^n_0 \over \partial \epsilon}=-\delta(\epsilon_n-\mu)$, and Eq.(\ref{cond_landsub}) simplifies to, 
\begin{eqnarray}
\sigma_{0l}=\frac{3}{4\pi^2}\frac{B}{B^*}\frac{E_{dc}}{E^*_{dc}}(\frac{k^3_F}{|\mu|^5}-\frac{k_F}{|\mu|^3})
\label{cond_T=0}
\end{eqnarray}
where $k_F=\sqrt{(\mu/\Delta)^2-(1-B(n+1/2)/B^*)^2}$ is the Fermi wave vector. It becomes zero for two magnetic fields: $B^\pm_n=(1\pm\mu/\Delta)B^*/(n+1/2)$, in $n$-th LL. The magnetic fields $B^\pm_n$ are real and finite for $\mu/\Delta<1$. For $\mu/\Delta=1$, the magnetic field $B^-_n$ is zero but $B^+_n$ has a real finite value. With $\mu/\Delta>1$, there is no real solution of  $B^-_n$ in the $k_F=0$ equation. This unique nature of $k_F$ is originated from the torus-like Fermi surface with two extremal surfaces $S_1$ and $S_2$ (shown in Fig.(\ref{fan_diag})). The energy dispersion for these extremal orbits (at $k_z=0$) reduces to $\epsilon_\pm=|\hbar^2(k^2_x+k^2_y)/2m_* -\Delta|$. With increasing values of $\mu$, the surface $S_2$ ($S_1$) diminishes (increases) and finally vanishes at $\mu=\Delta$. Consequently, there is only a single extremal surface $S_1$ remains for $\mu>\Delta$. The area of cyclotron orbit $A$ in $k$-space of $n$-th LL satisfies,
\begin{eqnarray}
\frac{\mathcal{A}\hbar}{e}\frac{1}{B}=2\pi(n+\gamma)
\label{cross-section}
\end{eqnarray}
where $0\leq \gamma<1$ is related to the Berry phase of that orbit. The frequency of quantum oscillation with the variation of $1/B$ is given by $F=\mathcal{A}_F\hbar/(2\pi e)$, where $\mathcal{A}_F$ is the cross section of extrimal Fermi surface. We find the oscillation frequencies are $F_{S_1}=B^*(1+\mu/\Delta)$ for outer circle and $F_{S_2}=B^*(1-\mu/\Delta)$ for inner circle, respectively. The finite temperature oscillation of NLSMs corresponds to the are $\mathcal{A}=0$ in $k$-space. However, the $1/B$ periodicity is governed by the area $\mathcal{A}_0=\pi p^2_0$.

In Fig.(\ref{low_temp_sigma}) we show the variation of second harmonic conductivity $\sigma(2\omega)$ as a function inverse magnetic field ($1/B$) $\mu>\Delta$ and $\mu<\Delta$. The variation of $\sigma(2\omega)$ with magnetic field is shown in the inset. The conductivity $\sigma(2\omega)$ with magnetic field exhibits oscillation where the amplitude increases with $B$. The oscillation occurs each time when a Landau level crosses the chemical potential, contributing a single peak in the conductivity. Each peak corresponds to the intraband transition in the Landau subbands. The higher Landau level contributes most to the $\sigma(2\omega)$ with decreasing magnetic field. As the magnetic field strength increases the lower Landau level starts to contribute. In the strong magnetic field regime when the Fermi level crosses lowest Landau level $n=0$, we have a cutoff value of magnetic field $B_c$. The Fermi wave vector $k_F$ becomes imiginary for any Landau subband $n$ when $B>B_c=2(1+\mu/\Delta)B^*$. Consequently, the conductivity $\sigma(2\omega)$ vanishes if $B>B_c$.  The periodicity of $\sigma(2\omega)$ with $1/B$ is given by, $(1/B)_{n+1}-(1/B)_{n}=1/[B^*(1+\mu/\Delta)]$. We find the oscillation period is $\approx .33/B^*$ for a given value of $\mu/\Delta=2$. On the other hand, for $\mu/\Delta<1$, the oscillation in $\sigma(2\omega)$ appears with two peaks with periodicity  $\Delta(1/B)=1/B^*(1-\mu/\Delta)$ and  $\Delta(1/B)=1/B^*(1+\mu/\Delta)$, respectively. We find oscillation frequencies $\Delta (1/B)=.88/B^*$ and $\Delta (1/B)=1.15/B^*$ for a given value of $\mu/\Delta=0.15$.

Fig.(\ref{fan_diag}) displays the Landau index plot, $n$ with peak position $B^*/B$ of SHC for three different values of $\mu$ ($\mu_3>\mu_2>\mu_1$). From Eq.(\ref{cross-section}), the Landau index $n$ is linearly proportional to $1/B$ for a given cross-sectional area $A$. The slope of the $1/B$ vs $n$-lines is proportional to the inverse of cross sectional area and the intercept of this line on the $n$-axis determines the value $-\gamma$. The value of $\gamma=0$ and $\gamma=1/2$ corresponds to Berry phase $0$ (trivial oscillation) and $\pi$ (topological oscillation)\cite{Oro-PRB18}, respectively. Fig.(\ref{fan_diag}) shows three fundamental frequencies. Two blue lines belong to the same frequency group, equal to $1/B^*(1-\mu/\Delta)$. Two red lines and black lines in Fig.(\ref{fan_diag}) belong to the other frequency group, equal to $1/B^*(1+\mu/\Delta)$. All the lines in Fig.(\ref{fan_diag}) show common intersections ($-1/2$) on the $n$-axis, indicating no Berry phase\cite{Rhim-PRB15,Deng-PRL19,Wang-PRL16,Kwan-PRR20,Zhijie-PRB17}.

We estimate the strength of current induced SHG by evaluating the nonlinear suscepibility $\chi(2\omega)=\sigma(2\omega)/2i\omega\epsilon_0$, where $\epsilon_0$ is the vauum permittivity. The susceptibility takes the largest value at the gap closing condition i.e., $B^*/B=(n+1/2)$. Considering, $m^*=0.1 m_e$ ($m_e$ is the electron's rest mass), $\Delta=0.01$ eV and $\tau=10^{-13}$s, we obtain $B^*\simeq 9$T and $E^*_{dc}\simeq 10^3$ V/m, respectively. We take the frequency of the probe light $\omega=0.5$ THz and the optical field $E_{op}=10^4$ V/m.  For a magnetic field value $B\simeq 6$T and $\mu=20$ meV, we obtain $j(2\omega)=\sigma(2\omega)E^2_{op}\simeq 10^{6}$ A/m$^2$ at low temperature $T=1K$ and $j(2\omega)\simeq 10^8$ A/m$^2$ at high temperature $T=150K$. We find $\chi(2\omega) \simeq 10^4$ pm/V and $\chi(2\omega)\simeq 10^6$ pm/V at temperature $T=1K$ and $T=150K$, respectively. These values of suscepibility are large compared to the value found in WSMs\cite{Wu-Nat17,Gao-PRB21}.

\textbf{Conclusions:-} We uncover an unusual mechanism for the realization of giant nonlinear response in NLSMs at finite temperature. This giant response occurs near specific values of magnetic fields for which Landau levels simplify to a linear spectrum. We find SdH oscillation in SHG which may probe the Fermi surface of NLSMs. We hope our findings will prompt further investigation of the nonlinear effect in NLSMs for technological applications.

\textbf{Acknowledgement:-} D.S. acknowledge T. K. Bose for initial collaboration.

\textbf{Note:-} During preparation of the manuscript we noticed Ref.\cite{Devakul} where similar oscillation are studied but not for nonlinear responses.

\end{document}